%% file: usenix.tex
\documentclass[conference]{IEEEtran}
\IEEEoverridecommandlockouts

\PassOptionsToPackage{table}{xcolor}  % belt-and-suspenders
\usepackage{xcolor}                   % loads with [table] via PassOptions

\usepackage{cite}
\usepackage{amsmath,amssymb,amsfonts}
\usepackage{algorithmic}
\usepackage{graphicx}
\usepackage{textcomp}
\usepackage{float}

\usepackage[most]{tcolorbox}
\usepackage{enumitem}
\usepackage{verbatim}
\usepackage{caption}
\usepackage{tabularray}
\usepackage{colortbl}
\usepackage{tikz}
\usetikzlibrary{trees}
\usepackage{url}
\usepackage{fancyhdr}

\def\BibTeX{{\rm B\kern-.05em{\sc i\kern-.025em b}\kern-.08em
    T\kern-.1667em\lower.7ex\hbox{E}\kern-.125emX}}
\begin{document}

\title{SIExVulTS: Sensitive Information Exposure Vulnerability Detection System using Transformer Models and Static Analysis\\
}

\author{
\IEEEauthorblockN{Kyler Katz\textsuperscript{1}, Sara Moshtari\textsuperscript{2}, Ibrahim Mujhid\textsuperscript{3}, Mehdi Mirakhorli\textsuperscript{4}, Derek Garcia\textsuperscript{5}}
\IEEEauthorblockA{
\textit{Department of Information and Computer Sciences} \\
\textit{University of Hawaii at Manoa}, United States \\
\textsuperscript{1}kkatz@hawaii.edu, 
\textsuperscript{2}saramsht@gmail.com, 
\textsuperscript{3}ijmujhid@hawaii.edu, 
\textsuperscript{4}mehdi23@hawaii.edu, 
\textsuperscript{5}dgarcia2@hawaii.edu
}
}

% \author{
% \IEEEauthorblockN{Kyler Katz}
% \IEEEauthorblockA{\textit{Department of Information and Computer Sciences} \\
% \textit{University of Hawaii at Manoa} \\
% United States \\
% kkatz@hawaii.edu}
% \and
% \IEEEauthorblockN{Sara Moshtari}
% \IEEEauthorblockA{\textit{Department of Information and Computer Sciences} \\
% \textit{University of Hawaii at Manoa} \\
% United States \\
% saramsht@gmail.com}
% \and
% \IEEEauthorblockN{Ibrahim Mujhid}
% \IEEEauthorblockA{\textit{Department of Information and Computer Sciences} \\
% \textit{University of Hawaii at Manoa} \\
% United States \\
% ijmujhid@hawaii.edu}
% \and
% \IEEEauthorblockN{Mehdi Mirakhorli}
% \IEEEauthorblockA{\textit{Department of Information and Computer Sciences} \\
% \textit{University of Hawaii at Manoa} \\
% United States \\
% mehdi23@hawaii.edu}
% \and
% \IEEEauthorblockN{Derek Garcia}
% \IEEEauthorblockA{\textit{Department of Information and Computer Sciences} \\
% \textit{University of Hawaii at Manoa} \\
% United States \\
% dgarcia2@hawaii.edu}
% }

\IEEEoverridecommandlockouts
\IEEEpubid{\makebox[\columnwidth]{\textnormal{979-8-3315-9147-2/25/\$31.00 \textcopyright2025 IEEE} \hfill}%
\hspace{\columnsep}\makebox[\columnwidth]{ }}

\maketitle

\input{Abstract}

\begin{IEEEkeywords}
Software Security, Static Analysis, Transformer Models, CWE-200, Sensitive Information Exposure

\end{IEEEkeywords}

\input{Introduction}

\IEEEpubidadjcol

\input{Motivation}

\input{RelatedWork}
\input{Methodology}

\input{Experiments}
\input{Discussion}
\input{Ethics}
\input{Conclusion}
\input{Acknowledgement}

{\footnotesize \bibliographystyle{acm}
\bibliography{references}}

\input{Appendix}

\end{document}

%% file: Abstract.tex
\begin{abstract}
\textit{Background:}
Sensitive Information Exposure (SIEx) vulnerabilities (CWE-200) remain a persistent and under-addressed threat across software systems, often leading to serious security breaches. Existing detection tools rarely target the diverse subcategories of CWE-200 or provide context-aware analysis of code-level data flows.

\textit{Aims:}
This paper aims to present SIExVulTS, a novel vulnerability detection system that integrates transformer-based models with static analysis to identify and verify sensitive information exposure in Java applications.

\textit{Method:}
SIExVulTS employs a three-stage architecture: (1) an Attack Surface Detection Engine that uses sentence embeddings to identify sensitive variables, strings, comments, and sinks; (2) an Exposure Analysis Engine that instantiates CodeQL queries aligned with the CWE-200 hierarchy; and (3) a Flow Verification Engine that leverages GraphCodeBERT to semantically validate source-to-sink flows. We evaluate SIExVulTS using three curated datasets, including real-world CVEs, a benchmark set of synthetic CWE-200 examples, and labeled flows from 31 open-source projects.

\textit{Results:}
The Attack Surface Detection Engine achieved an average F1 score greater than 93\%, the Exposure Analysis Engine achieved an F1 score of 85.71\%, and the Flow Verification Engine increased precision from 22.61\% to 87.23\%. Moreover, SIExVulTS successfully uncovered six previously unknown CVEs in major Apache projects.

\textit{Conclusions:}
The results demonstrate that SIExVulTS is effective and practical for improving software security against sensitive data exposure, addressing limitations of existing tools in detecting and verifying CWE-200 vulnerabilities.
\end{abstract}

%% file: Introduction.tex
\section{Introduction}
 Software systems are critical components of modern infrastructure. As these systems grow in size and complexity, they become increasingly susceptible to various vulnerabilities. One primary concern when assessing the security of software systems is the exposure of sensitive and critical information, such as user credentials, to unauthorized users. Detecting and addressing this type of vulnerability in the early stages of software development can lead to enhanced security, a reduced risk of data breaches, lower costs associated with fixing vulnerabilities post-release, and improved user trust and compliance with data protection regulations. 

The exposure of sensitive information to unauthorized actors \cite{CWE200} can be an initial step in an attack, providing valuable insights into the system that enable further malicious actions. Therefore, implementing a robust strategy to detect and mitigate this vulnerability is crucial. For instance, if the location of a configuration file is disclosed and accessible, an attacker might obtain credentials for accessing the database, leading to more severe security breaches.

Prior works in vulnerability detection mainly focused on detecting different types of vulnerabilities \cite{li2018vuldeepecker,lin2018cross} and evaluated the general performance of the proposed vulnerability detection models. Some other studies focused on detecting specific types of vulnerabilities such as \textit{SQL Injection} \cite{TANG2020105528}, \textit{Cross-site Scripting}\cite{fang2018deepxss}, and memory-related vulnerabilities\cite{cao2023learning,wang2021spotting}. Despite the importance of the \textit{Sensitive Information Exposure (SIEx)} vulnerabilities \cite{CWE200,iex1},  investigation and detection of different types of SIEx vulnerabilities has not been studied in the literature.
The exposure of sensitive information encompasses various types of vulnerabilities. The Common Weakness Enumeration (CWE) repository \cite{CWE} provides a hierarchical model for these types of vulnerabilities under the CWE-200 category. This hierarchical model presents that CWE-200 includes different subcategories that describe different kinds of SIEx vulnerabilities. Detection of SIEx vulnerabilities requires an extensive analysis of the CWE-200 subcategories and an understanding of, and approach to, detecting vulnerabilities in this category.

In this study, we utilize a hierarchical model based on CWE-200 and employ a combination of transformer models and static analysis to detect SIEx vulnerabilities. By examining the CWE-200 hierarchical model, along with code examples and additional information provided in the CWE catalog, we identify attack surface components related to these vulnerabilities. 

Attack surface components are critical parts of software that contain valuable content or are targeted by attackers to gain system access or perform malicious actions \cite{moshtari2022grounded}. For CWE-200, sensitive data (sources) and potential points of data exposure (sinks) are considered attack surface components. Some vulnerabilities related to CWE-200 focus on the types of sensitive data exposed to attackers \cite{CWE359}. In contrast, others describe specific parts of an application where sensitive information can be exposed to unauthorized users, such as through runtime errors that print sensitive information \cite{CWE357}.

\begin{figure*}[t!]
  \centering
  \resizebox{\textwidth}{!}{
    \begin{tikzpicture}[
      sibling distance=9em,
      level distance=5em,
      every node/.style = {
        draw,
        align=center,
        font=\scriptsize,
        top color=white,
        bottom color=blue!10,
        rectangle,
        rounded corners
      }
    ]
    \node {\shortstack{CWE-200:\\Exposure of Sensitive\\Information to an Unauthorized Actor}}
      child { node {\shortstack{CWE-201:\\Insertion of Sensitive\\Information Into Sent Data}}
        child { node {\shortstack{CWE-598:\\GET Request with\\Sensitive Query Strings}} }
      }
      child { node {\shortstack{CWE-538:\\Sensitive Info in\\Accessible File/Directory}}
        child { node {\shortstack{CWE-532:\\Sensitive Info in\\Log File}} }
      }
      child { node {\shortstack{CWE-214:\\Sensitive Info in\\Process Invocation}} }
      child { node {\shortstack{CWE-203:\\Observable Discrepancy}}
        child { node {\shortstack{CWE-204:\\Observable Response\\Discrepancy}} }
        child { node {\shortstack{CWE-208:\\Observable Timing\\Discrepancy}} }
      }
      child { node {\shortstack{CWE-215:\\Sensitive Info in\\Debugging Code}} }
      child { node {\shortstack{CWE-615:\\Sensitive Info in\\Source Code Comments}} }
      child { node {\shortstack{CWE-209:\\Error Message Containing\\Sensitive Information}}
        child { node {\shortstack{CWE-535:\\Shell Error Message}} }
        child { node {\shortstack{CWE-536:\\Servlet Runtime Error\\Message with Sensitive Info}} }
        child { node {\shortstack{CWE-537:\\Java Runtime Error\\Message with Sensitive Info}} }
        child { node {\shortstack{CWE-550:\\Server-Generated Error\\Message with Sensitive Info}} }
      };
    \end{tikzpicture}
  }
  \caption{CWE-200 Hierarchical Model}
  \label{fig:cwe200-tree}
\end{figure*}
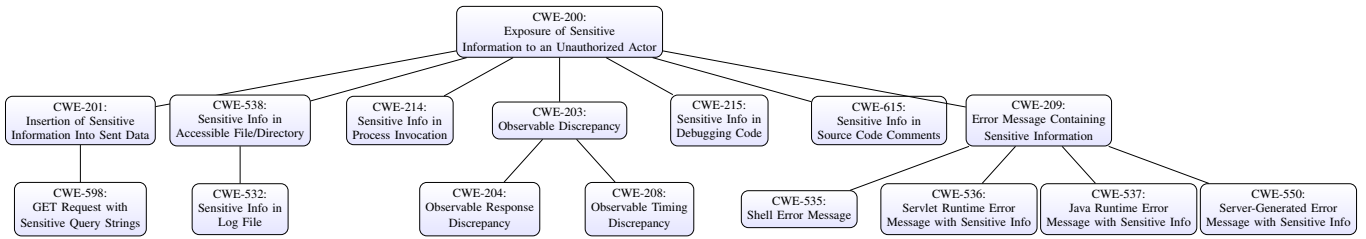

In summary, our contributions are as follows:
\begin{itemize}
\item We present a three-stage system for detecting CWE-200 vulnerabilities: (1) a transformer-based Attack Surface Detection Engine to identify sensitive elements and sinks, (2) a CodeQL-driven Exposure Analysis Engine to trace flows between them, and (3) a Flow Verification Engine that uses code context, dataflow, and transformer embeddings to reduce false positives.
\item We define a comprehensive categorization of sensitive information types in source code by analyzing CWE descriptions, real-world vulnerabilities from the NVD, and common coding patterns.
\item We introduce three labeled datasets—including benchmark (synthetic examples), CVE (CVE-derived code), and SIEx-Flow Dataset (dataflows) from 31 open-source projects—that support sensitive information detection and dataflow-based verification.
\item We benchmark transformer models—including ChatGPT-4o, SentBERT, CodeBERT, and CodeT5—for their ability to detect sensitive elements in source code, selecting the most effective model for integration into SIExVulTS.
\item Experimental results demonstrate that SIExVulTS detects a wide range of CWE-200 vulnerabilities with high accuracy on both benchmark and real-world projects.
\end{itemize}

% In this paper, we aim to develop a detection system for Sensitive Information Exposure (SIEx) vulnerabilities that leverages transformer models and static analysis to accurately identify and verify sensitive data exposure. To this end, we introduce SIExVuTS, a three-stage vulnerability detection system that combines transformer-based classification, CodeQL-driven static analysis, and semantic flow verification to detect CWE-200 vulnerabilities with high precision. Our approach is evaluated on a mix of benchmark and real-world datasets, including newly discovered CVEs in widely used open-source projects.

The remainder of this paper is organized as follows: Section~\ref{sec:motiv} describes the CWE-200 vulnerability and provides a motivating example.  Related studies are described in Section~\ref{sec:relatedwork}. Section~\ref{sec:methodology} provides an overview of the methodology used in this study. Section~\ref{sec:experiments} presents the collected data and experiments that have been performed to evaluate the SIExVulTS tool alongside existing tools. Section~\ref{sec:discussion} discusses the results and threats to validity. Finally, Section~\ref{sec:conclusions} concludes the paper.

%% file: Motivation.tex
\section{Background }
\label{sec:motiv}
\subsection{CWE-200 Vulnerabilities}

CWE-200 vulnerabilities occur when a software system exposes sensitive information to unauthorized users. Sensitive information, such as passwords, credit card numbers, health records, and other forms of personal information (PII), constitutes critical attack surface components that attackers can target \cite{moshtari2022grounded}. 
According to OWASP, in 2021, \textit{Sensitive Data Exposure} was number two on their \textbf{Top 10 Web Application Security Risks} \cite{iex1}. 

Despite the critical nature of this vulnerability, there is no comprehensive research dedicated explicitly to detecting these vulnerabilities. Most existing vulnerability detection approaches focus on identifying various types of vulnerabilities, but do not explicitly cover CWE-200 or its sub-categories. By detecting these vulnerabilities and providing detailed information to developers, it becomes easier to address and mitigate these security weaknesses in source code.

\subsection{CWE-200 Hierarchical Model}

To address this vulnerability and its sub-categories, we use a hierarchical model as seen in Fig.\ref{fig:cwe200-tree} inspired by The Common Weakness Enumeration (CWE) repository  \cite{CWE} and adapted to suit the needs and limitations of our tool.

\begin{figure*}[h]
    \centering
        \includegraphics[width=1\linewidth]{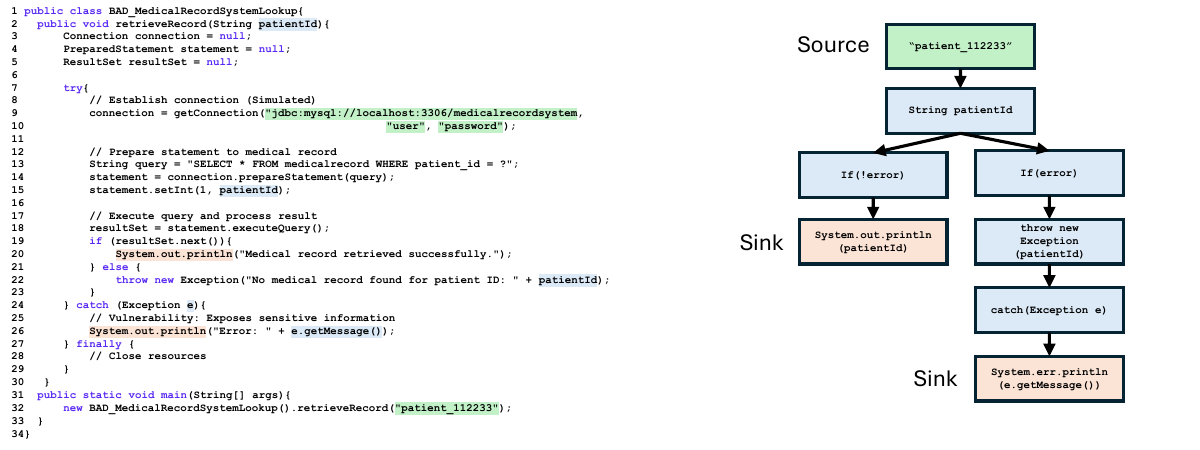}
    \caption{Example illustrating the data flow from source to sink, demonstrating how sensitive information propagates through the system}
    \label{fig:example}
\end{figure*}

\subsection{Sensitive Data} 
Sensitive data is categorized into eight types based on analysis of CWEs and NVD reports:
\begin{enumerate}
\item \textit{Authentication, Authorization, and Credentials Information}: Data holding passwords, API keys, usernames, verification codes, credentials, etc.
\item \textit{Personal Identifiable Information (PII)}: Sensitive data containing names, emails, addresses, social security numbers, health information, accounts, etc.
\item \textit{Financial Information}: Variables related to credit cards, bank account numbers, account IDs, payment IDs, CVV, etc.
\item \textit{Files Containing Sensitive Information, Sensitive File Paths, URLs/URIs}: Variables storing internal URLs/URIs or file paths to files that contain sensitive information or files themselves.
\item \textit{Sensitive System and Configuration Information}: Data with database, cloud provider, or network connection strings, database schemas, configuration details, environment variables, sensitive settings, controllers, and managers.
\item \textit{Security and Encryption Information}: Data holding encryption keys, seeds, or certificates.
\item \textit{Application-Specific Sensitive Data}: Data storing sensitive information such as device details (names, IDs, properties, objects), application-specific IDs, email messages, notifications, etc.
\item \textit{Query Parameters}: Variables storing sensitive data in HTTP GET requests.
\end{enumerate}

\subsection{Motivating Example}
The example in Fig. \ref{fig:example} provides a code snippet that demonstrates a classic example of CWE-537, where a medical record retrieval system inadvertently exposes sensitive patient information due to improper exception-handling practices. The code includes sensitive information, such as "\textit{patient\_112233}" (line 32), the database connection URL, username, and password (lines 9 and 10). The "\textit{patient\_112233}" is sent to the "\textit{retrieveRecord(String patientId)}" method as an input parameter. Specifically, the code prints the patient ID if a medical record cannot be retrieved (line 18). It includes it in exception error messages when no record is found or when an error occurs. This practice can lead to unauthorized access to sensitive information, potentially violating privacy regulations and leading to significant security breaches. Exposure of such information can be exploited by malicious actors to gain further unauthorized access or to perpetrate identity theft. By addressing these vulnerabilities, the application can better protect sensitive medical data and comply with privacy standards, thus maintaining the confidentiality and integrity of patient information.

%% file: RelatedWork.tex
\section{Related Work}
\label{sec:relatedwork}
In this section, we describe the related work of three aspects:
\subsection{Large Language Models (LLM)}
LLMs have been widely applied to natural language tasks (e.x. sentiment analysis \cite{kheiri2023sentimentgpt}, translation \cite{brown2020language}, etc.), and they are also getting more popular among software practitioners for different \textit{Software Engineering} (\textit{SE}) tasks. They have been used in code generation \cite{koziolek2024llm}, test case generation \cite{kang2023large,wang2024efficiently}, program repair \cite{xia2023automated}, requirement completeness \cite{luitel2024improving}, resolving code quality issues \cite{wadhwa2024core}, etc. Leveraging ChatGPT to perform sub-tasks \cite{chatgpt} has been the focus of many studies. Jalil et al. \cite{jalil2023chatgpt} investigated the effectiveness of ChatGPT in answering test questions, while Ahmad et al. \cite{ahmad2023towards} explored its application in system architecture. Sun et al. \cite{sun2023automatic} discussed the challenges of using ChatGPT for code summarization. In addition to addressing textual problems in SE, other studies have evaluated ChatGPT's performance in code-related tasks such as program repair \cite{cao2023study} and code generation \cite{liu2024no}.
\subsection{Vulnerability Analysis}
Recent studies in software vulnerability detection mostly focused on general vulnerability detection \cite{moshtari2016evaluating,chakraborty2021deep,cheng2021deepwukong}. These studies extract \textit{Abstract Syntax Trees} \cite{lin2018cross} or program slices by considering source and sink \cite{li2018vuldeepecker} and provide deep learning-based models to detect which program slice is vulnerable and which one is non-vulnerable. There are a few studies in the literature that focus on identifying vulnerable code components based on CWE types \cite{cheng2021deepwukong}. However, these studies do not cover CWE-200. Pan et al. \cite{pan2023fine} proposed a model to classify vulnerability types to make it easier for practitioners to analyze vulnerabilities. In addition to general vulnerability detection, literature also focuses on the detection of specific vulnerability types such as \textit{SQL Injection} \cite{TANG2020105528}, \textit{Cross-Site Scripting} \cite{fang2018deepxss}, and memory-related vulnerabilities \cite{cao2023learning} such as \textit{Buffer Overflow}\cite{wang2021spotting}.
\subsection{Attack Surface Detection}
To detect \textit{source} and \textit{sink} for program analysis, most studies rely on predefined lists of API/function calls. Li et al. \cite{li2018vuldeepecker} considered user input as the source and extracted 56,902 library/API function calls from programs, including library/API function calls related to buffer and resource management errors. Bian et al. \cite{bian2020sinkfinder} proposed a machine learning-based approach named \textit{SinkFinder} to detect these vital components in source code. Huang et al. \cite{huang2024raisin} proposed a context-based approach to detect rare sensitive functions in source code.

%% file: Methodology.tex
\section{Method}
\label{sec:methodology}
Our objective is to design a vulnerability detection system that automatically identifies CWE-200 vulnerabilities and their precise locations in source code.

This is achieved through a multi-stage process that detects sensitive data (sources), identifies exposure points sinks), analyzes dataflows between them, and verifies the validity of the identified paths. The system leverages transformer-based models, static analysis via CodeQL, and a dedicated flow verification step to ensure accuracy.

\begin{figure*}[t]
    \centering
    \includegraphics[width=\linewidth]{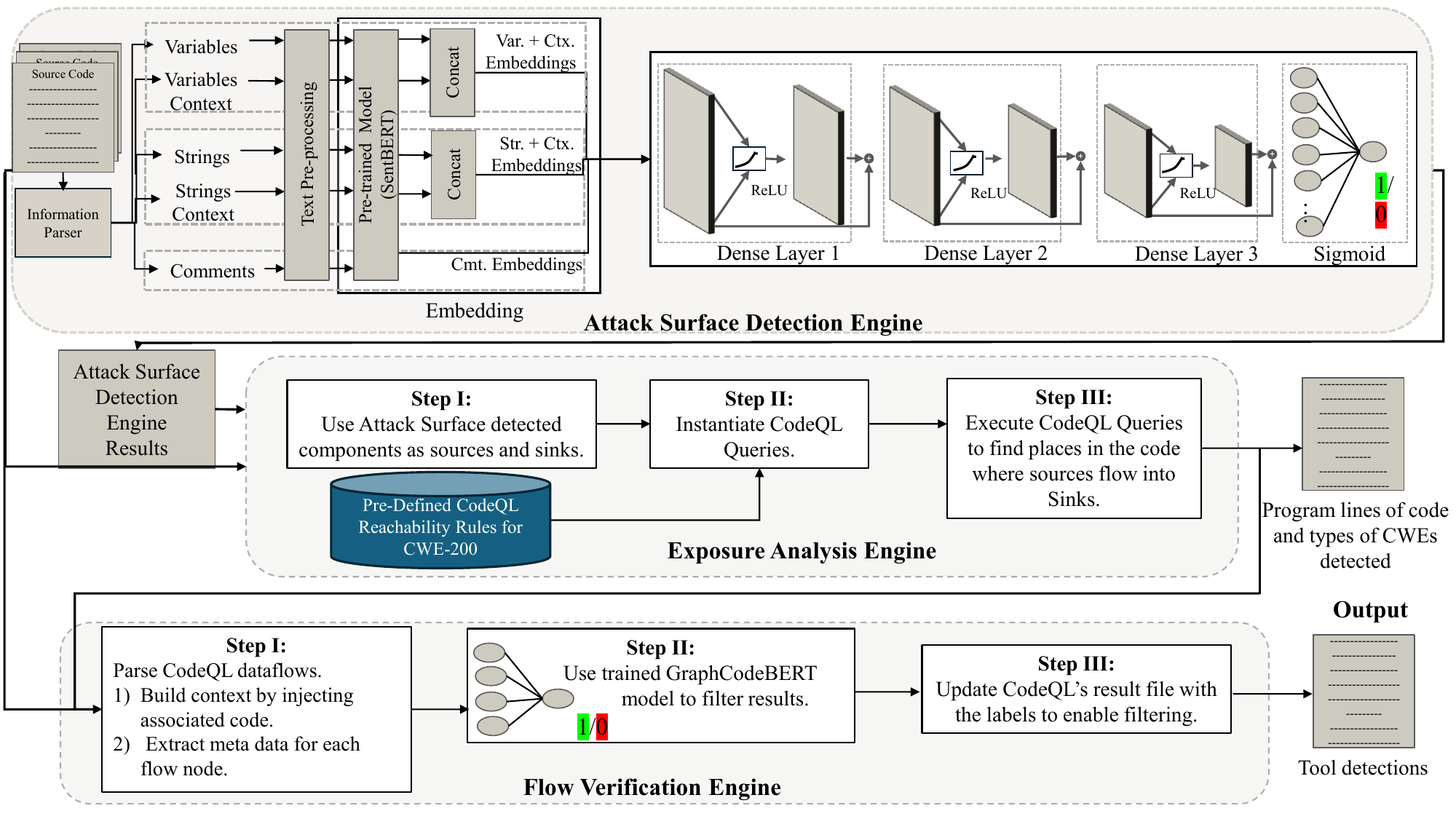}
    % \captionsetup{labelformat=empty}
    \caption{An overview of the SIExVulTS tool.
 }
    \label{fig:model}
\end{figure*}

As illustrated in Fig.~\ref {fig:model}, the architecture of SIExVulTS consists of three major components: the \textit{Attack Surface Detection Engine}, the \textit{Exposure Analysis Engine}, and the \textit{Flow Verification Engine}. These components operate in sequence to identify attack surface elements, trace dataflows, and verify actual vulnerability exposures aligned with the CWE-200 hierarchy.

\subsection{Attack Surface Detection Engine} \subsubsection{Detecting Sensitive Information} 
Identifying sensitive information within codebases has traditionally relied on scanning for commonly recognized sensitive terms \cite{Yang2018} such as "cardNumber," "password," or "APIkey." Although this method is straightforward to implement, its effectiveness is constrained by the scope of its predefined lists. These lists may not encompass all instances of sensitive data, especially rare or unseen sensitive data that is unique to specific projects or contexts. 

To overcome these limitations, we leverage transformers to identify sensitive data in source code. These advanced models enhance our detection capabilities by taking into account data context. This contextual awareness enables the model to identify sensitive elements that traditional methods might not detect, providing a more robust solution for securing source code. For instance, in the case of an Android source file associated with Google Pay, the model assesses potential information leaks within the specific framework of payment processing. 

We evaluate multiple fine-tuned transformer-based models, including ChatGPT-4o \cite{chatgpt}, SentBERT \cite{reimers2019sentence}, CodeBERT \cite{feng2020codebertpretrainedmodelprogramming}, and CodeT5 \cite{raffel2023exploringlimitstransferlearning}. ChatGPT-4o with zero-shot is used as a baseline.

\textbf{Encoder-Based Detection Approach:} In this approach, we use embeddings from SentBERT, CodeBERT, and CodeT5 to detect sensitive data. Each model was chosen for its strengths:
\begin{itemize}
\item \textbf{SentBERT} captures contextual meaning in natural language and is effective for identifying nuanced elements like comments or ambiguous variable names.
\item \textbf{CodeBERT} trained on code and natural language pairs, excels at capturing code structure and semantics, especially for identifying APIs and function-related elements.
\item \textbf{CodeT5} is a code-specific encoder-decoder model trained on code and natural language, and performs well on mixed-format inputs such as source code with inline comments or documentation.
\end{itemize}

Two types of embeddings from each of these models are utilized to classify data as sensitive or non-sensitive, name embeddings and context embeddings.

\begin{enumerate}
\item \textbf{Name Embeddings}: These embeddings represent the literal value of the data in question. For a variable, this corresponds to the variable name; for a string, it is the string's value; for a comment, the comment text; and for an API call, the name of the call.

\item \textbf{Context Embeddings}: These embeddings capture the context in which the data appears, incorporating additional details such as the data type, where applicable. Multiple iterations were tested during development to determine the optimal context representation:
\begin{itemize}
\item File-Level Context: Embedding the entire file introduced excessive noise, leading to poor classification performance.
\item Line-Level Context: Limiting context to only the lines where the data appears proved too narrow, resulting in frequent misclassifications.
\item Method/Line Hybrid Context: The final approach embeds the entire method where the data appears, along with any relevant lines in the global scope. Additionally, the data type of the variable or element is included to enhance the contextual representation. This approach balances sufficient contextual information while minimizing noise. 
\end{itemize}
\end{enumerate}

We use the hybrid approach as it provides the best context while balancing noise. 

As shown in Fig. \ref{fig:model}, we pre-processed variables, strings, comments, and sinks by extracting both their names and relevant code context (excluding comments, which are contextually self-contained). These inputs are embedded using one of the transformer models, producing vector representations of both the raw identifier and the surrounding code. For variables, strings, and sinks, name and context embeddings are concatenated.

The combined vectors serve as input to either a binary classifier (for identifying sensitive elements) or a multi-class classifier (for categorizing sink types). The binary classifier uses a four-layer architecture with skip connections and residual layers for deeper feature learning. At the same time, the sink model follows a similar design with categorical softmax output over eight sink types.

\subsubsection{Training and Hyperparameter Optimization}

Each neural classifier was trained using the embeddings described above, passed through a four-layer fully connected architecture. The first dense layer was initialized to one-third of the input size, with each subsequent layer halved in width. Binary classifiers (for variables, strings, and comments) used a single sigmoid output neuron, while the sink classifier used a softmax layer with eight output classes.

To prevent overfitting, L2 regularization and dropout were applied to all hidden layers. Hyperparameters were optimized via a randomized grid search over 50 trials, using a fixed random seed for reproducibility. The search explored the following space:
\begin{itemize}
    \item \textbf{Learning rate:} \{1e-5, 1e-4, 1e-3, 1e-2\}
    \item \textbf{Dropout rate:} \{0.2, 0.3\}
    \item \textbf{Activation functions:} ReLU, ELU, Sigmoid
    \item \textbf{Batch size:} \{32, 64\}
    \item \textbf{Epochs:} \{50, 60\}
\end{itemize}

A stratified 70:15:15 train/validation/test split was used to ensure representation across classes. During training, early stopping monitored validation loss with a patience of 10 epochs, and learning rate reduction was triggered when the loss plateaued. Final model selection was based on the highest F1-score on the validation set.

To further enhance robustness and mitigate data variance, 5-fold stratified cross-validation was employed during hyperparameter tuning. SMOTE \cite{Chawla_2002} was applied to all binary tasks to address class imbalance.

\textbf{Rationale:} These choices were made to balance expressiveness and generalizability while ensuring efficient training. Adam was selected as the optimizer for its adaptive gradient scaling. F1-score was used as the primary metric due to its sensitivity to class imbalance—especially critical when identifying rare but high-risk sensitive elements and sinks.

\subsection{Exposure Analysis Engine}
The exposure analysis engine, as shown in Fig. \ref{fig:model}, utilizes CodeQL\cite{codeql}, a robust and efficient static analysis tool that transforms a codebase into a queryable database. This enables the execution of complex queries to detect dependencies and interactions within the code. To maximize the potential of CodeQL, we developed a reusable set of queries aligned with our CWE-200 hierarchical model. This process involved a detailed review of CWEs as outlined in the CWE repository \cite{CWE} and the creation of targeted queries to effectively identify CWE-200 vulnerabilities. A CodeQL query is a declarative script designed to detect specific patterns, bugs, or codebase issues by leveraging data flow analysis. Using these queries, we defined logic to capture the flow of sensitive information from sources to sinks, identifying potential exposures for each CWE. 

For example, CWE-537, which addresses the insertion of sensitive information into runtime error messages, requires identifying variables containing sensitive data (sources) and tracing their flow to exception-handling operations. The query checks whether sensitive information is exposed while handling a caught exception, ensuring precise vulnerability detection. The exposure analysis engine is designed not only to detect vulnerabilities but also to provide detailed insights into the pathways through which sensitive information propagates. After identifying sensitive data sources and their corresponding sinks using our CodeQL queries, the engine outputs a comprehensive report highlighting The specific paths taken by sensitive information from source to sink. Along with the contexts in which the exposure occurs, such as exception handling, API calls, or file operations.

\subsection{Flow Verification Engine}
The Flow Verification Engine enhances CodeQL-generated dataflows to assess whether each source-to-sink path reflects a true vulnerability or a false positive. Although CodeQL provides dataflow node locations, these are often abstract and lack the surrounding source code necessary for contextual understanding. As a first step, the engine reconstructs and enriches each flow by parsing the original source files to extract and inject relevant code surrounding each node. Each step in the flow is labeled with its corresponding code snippet, data type, and semantic role, forming a more comprehensive and interpretable path representation.

Following this augmentation, the engine deduplicates flows using SHA-256 hashing to eliminate repeated instances. Each unique flow is then serialized into a string that includes the CWE identifier and the full propagation path. These enriched flows are embedded using GraphCodeBERT, a transformer model designed for code semantics. To handle long sequences, a Transformer-based aggregator compresses token-level embeddings from segmented input, enabling representation of deeply nested objects and extended propagation chains.

The resulting embeddings are used to train a residual neural classifier, similar in architecture to that of the \textit{Attack Surface Detection Engine}. Training is conducted using stratified splits, SMOTE for balancing, and randomized hyperparameter search. This engine significantly boosts system precision by filtering false positives that static analysis alone may misclassify.

\subsection{Summary}
SIExVulTS integrates its three core components into a cohesive detection pipeline. The Attack Surface Detection Engine identifies sensitive sources and potential sinks. These components are then used to instantiate predefined CodeQL queries in the Exposure Analysis Engine, which traces possible dataflows between them. Finally, the Flow Verification Engine evaluates these paths for actual exposure, filtering out false positives and increasing precision.

The integration between these stages is realized through a structured three-step process:
\begin{itemize}
\item \textbf{Step I}: The Attack Surface Detection Engine analyzes the source code to identify attack surface components (sources and sinks). 
\item \textbf{Step II}: These elements are used to instantiate customized CodeQL queries for the given application, performing lightweight reachability analysis to determine whether any static path exists between detected sources and sinks. If a path is found, the Exposure Analysis Engine returns a dataflow trace, including the CWE identifier, source, sink, and the steps between them.
\item \textbf{Step III}: These flows are then passed to the Flow Verification Engine, which analyzes their semantic validity and likelihood of exposing sensitive data.
\end{itemize}

This staged process allows SIExVulTS to move from broad detection to precise, validated vulnerability identification, reducing noise and enabling accurate detection of CWE-200 vulnerabilities in diverse codebases.

% \begin{figure}[h]

%     \centering
%    % \includegraphics[width=\linewidth]{IEEE-conference-template-062824/images/Paper Code Example.pdf}
%         \includegraphics[width=0.7\linewidth]
%         {Conference-LaTeX-template_10-17-19/images/codesampledfg2.pdf}
%     % \captionsetup{labelformat=empty}
  
%     \caption{Data Flow Graph illustrating the data flow from source to sinks}
%     \label{fig:examplegraph}
    
% %     \label{fig:example}
% \end{figure}

%% file: Experiments.tex
\section{Results}
\label{sec:experiments}

\begin{table}[h!] % Use table* for spanning both columns
\centering % Centers the entire table within the page
\caption{Data statistic and number and percentage of attack surface components detected in each dataset\label{tbl:datastatistic}}
\begin{tblr}{
  width=\textwidth, % Ensure the table spans the full width of the page
  column{1} = {},
  column{2} = {},
  column{3} = {},
  column{4} = {},
  column{5} = {},
  column{6} = {},
  column{7} = {},
  cell{2}{1} = {r=2}{},
  cell{2}{7} = {r=2}{},
  cell{4}{1} = {r=2}{},
  cell{4}{7} = {r=2}{},
  cell{6}{1} = {r=2}{},
  cell{6}{7} = {r=2}{},
  cell{8}{1} = {r=2}{},
  cell{8}{7} = {r=2}{},
  vlines,
  hline{1-2,4,6,8,10} = {-}{},
  hline{3,5,7,9} = {2-6}{},
}
\textbf{Type} & \textbf{Label}         & \textbf{CVE} & \textbf{Bench} & \textbf{Total} & \textbf{\%} & \textbf{Sum} \\
% \hline
Variables & Sens     & 1138         & 389         & 1527  & 0.57       & 2683           \\
          & Non-Sens & 487          & 669         & 1156  & 0.43       &                \\
Strings   & Sens     & 316          & 20          & 336   & 0.14       & 2349           \\
          & Non-Sens & 1024         & 989         & 2013  & 0.86       &                \\
Comments  & Sens     & 11           & 4           & 15    & 0.01       & 1003           \\
          & Non-Sens & 385          & 603         & 988   & 0.99       &                \\
API Calls     & Sink     &       210       &  505           &    715   &       0.21     &   3443             \\
          & Non-Sink &     1636         &  1092          &   2728    &     0.79       &                
\end{tblr}
\end{table}

\begin{table*}[t]
\caption{Weighed performance comparison of GPT-4o, SentBERT, CodeBERT, and CodeT5 on identifying attack surface components (P = Precision, R = Recall, F = F1 Score).}
\centering
\begin{tabular}{|l|c c c|c c c|c c c|c c c|}
\hline
\textbf{Data Item} & \multicolumn{3}{c|}{\textbf{GPT-4o}} & \multicolumn{3}{c|}{\textbf{SentBERT}} & \multicolumn{3}{c|}{\textbf{CodeBERT}} & \multicolumn{3}{c|}{\textbf{CodeT5}} \\
\hline
& P & R & F & P & R & F & P & R & F & P & R & F \\
\hline
Variables 
& 0.7934 & 0.6142 & 0.6853 
& \cellcolor{gray!20}\textbf{0.9383} & \cellcolor{gray!20}\textbf{0.9387} & \cellcolor{gray!20}\textbf{0.9385} 
& 0.9223 & 0.9252 & 0.9226 
& 0.9282 & 0.9301 & 0.9259 \\
\hline
Strings 
& 0.3726 & 0.5318 & 0.4289 
& 0.9567 & 0.9548 & 0.9556 
& 0.9645 & 0.9651 & 0.9627 
& \cellcolor{gray!20}\textbf{0.9596} & \cellcolor{gray!20}\textbf{0.9610} & \cellcolor{gray!20}\textbf{0.9600} \\
\hline
Comments 
& 0.1283 & 0.2227 & 0.1624 
& 0.9776 & 0.9778 & 0.9777 
& 0.9932 & 0.9932 & 0.9931 
& \cellcolor{gray!20}\textbf{0.9983} & \cellcolor{gray!20}\textbf{0.9983} & \cellcolor{gray!20}\textbf{0.9983} \\
\hline
Sinks 
& 0.6889 & 0.6745 & 0.6718 
& \cellcolor{gray!20}\textbf{0.9768} & \cellcolor{gray!20}\textbf{0.9774} & \cellcolor{gray!20}\textbf{0.9763} 
& 0.9677 & 0.9688 & 0.9659 
& 0.9730 & 0.9731 & 0.9726 \\
\hline
\end{tabular}
\label{tbl:performance_comparison}
\end{table*}

We evaluate SIExVulTS using a series of experiments designed to assess its performance in identifying sensitive information exposure (SIEx) vulnerabilities. Specifically, we examine the effectiveness of the Attack Surface Detection Engine, the accuracy of the Exposure Analysis Engine, and the practical impact of the Flow Verification Engine on real-world projects. These evaluations are structured around three guiding research questions:

\begin{itemize}
\item \textbf{RQ1}: Which model most accurately identifies attack surface components?
\item \textbf{RQ2}: How effectively does SIExVulTS detect CWE-200 vulnerabilities in benchmark scenarios?
\item \textbf{RQ3}: Can SIExVulTS detect CWE-200 vulnerabilities in real-world open-source projects?
\end{itemize}

\subsection{Dataset Overview}
We validate our approach using two primary datasets and one supplemental dataset:

\subsubsection{CVE Dataset}
This dataset was curated from the National Vulnerability Database (NVD) and relevant code repositories. We selected 40 CVEs associated with CWE-200 vulnerabilities in open-source projects, analyzing their reported patches and extracting data items from pre-fix files. Each file was parsed to extract variables, hard-coded strings, comments, and API calls. These elements were then labeled as either sensitive/non-sensitive (or sink/non-sink in the case of API calls).

This dataset provides a real-world basis for evaluating the performance of the Attack Surface Detection Engine.

% \usepackage{tabularray}

%--------------------------------------------------
\subsubsection{Benchmark Dataset}
To ensure structured and reproducible evaluation of CWE-200 detection, we developed a benchmark dataset consisting of synthetic but representative code samples. Each sample is aligned with a specific CWE and categorized as either GOOD'' (no vulnerability) or BAD'' (contains vulnerability). These examples reflect common patterns observed in real-world vulnerabilities.

This dataset enables detailed testing of CodeQL-based detections. In total, it contains 300 labeled samples: 161 BAD and 139 GOOD.

%--------------------------------------------------

\subsubsection{Flow Verification Dataset} This dataset includes labeled dataflows from 31 real-world open-source Java projects, extracted using our predefined CodeQL queries. Each flow is manually labeled as \textit{Yes} (true positive) or \textit{No} (false positive) based on whether it results in actual sensitive data exposure. The dataset is used to train and evaluate the Flow Verification Engine, helping it learn to distinguish between valid and invalid flows. There are 2,555 data flows in total, varying in complexity from just a few steps to over 150, capturing the nuanced behaviors of information propagation in practical settings. 

\subsubsection{Additional Training Data}

As shown in Table~\ref{tbl:datastatistic}, sensitive strings and comments comprise only 14\% and 1\% of the collected data, respectively, posing a challenge for training the Attack Surface Detection models. To address this imbalance, we added synthetic strings and comments that reflect patterns observed in real vulnerabilities and align with our eight predefined categories of sensitive information. Specifically, we added 1,427 comments (690 sensitive and 737 non-sensitive) and 453 strings (69 sensitive and 384 non-sensitive) to enhance model training.

% OpenMRS: 2.6.7
% Jenkins: 2.107.1
% ClickHouse: 0.4.5
% Apache Tomcat:  7.0.16

% The first [TO BE FILLED OUT BY THE REST OF THE TEAM]
\subsection{Experimental Results}
% #-------------------------------------------------
% #-------------------------------------------------

% #-------------------------------------------------
% #-------------------------------------------------
\subsubsection{RQ1: Which model most accurately identifies attack surface components?}

We evaluated four models: ChatGPT-4o, SentBERT, CodeBERT, and CodeT5. Table~\ref{tbl:performance_comparison} summarizes their precision (P), recall (R), and F1 score (F) across the four attack surface categories.

\textbf{CodeT5} achieved the highest F1 scores for both strings and comments. This can be attributed to its encoder-decoder architecture, which is particularly well-suited for handling free-form text and mixed-format inputs. Its ability to model sequential dependencies likely helped it capture nuanced patterns within comments and hardcoded string literals—often areas where meaning is implicit or loosely structured.

\textbf{SentBERT}, while slightly behind CodeT5 in some categories, showed the most consistent performance overall. Its strength lies in generating compact semantic embeddings that are highly effective for classification tasks, particularly where clear contextual meaning can be inferred from code and identifier names. Notably, it excelled at variables and sinks, where sensitivity is often contextually tied to naming conventions and surrounding logic. SentBERT's 384-dimensional embeddings also offer a significant efficiency advantage over CodeT5's 768-dimensional ones, especially when processing large codebases with millions of elements.

\textbf{CodeBERT} performed well on Sink detection, likely due to its joint training on natural language and source code, which allows it to better understand function signatures and common coding idioms.

\textbf{ChatGPT-4o} was evaluated using a zero-shot prompting strategy, as few-shot prompting was experimentally found to degrade performance. While few-shot examples are typically used to guide model behavior, in this case, they introduced unintended biases and increased ambiguity, especially when the examples did not perfectly match the project-specific context of the inputs. As a result, the model often overfit to patterns in the examples rather than generalizing effectively. In contrast, zero-shot prompting—despite its simplicity—offered more consistent performance, though it still underperformed relative to encoder-based models. Its primary weaknesses included inconsistent handling of nuanced distinctions (e.g., between strings and comments) and susceptibility to prompt misinterpretation, especially when the surrounding context exceeded token limits. An example can be found in Fig. \ref{fig:prompt}.

Based on these observations, SIExVulTS uses SentBERT as its primary embedding model due to its competitive performance across all categories, efficient embedding size, and strong generalization in both structured and unstructured code elements. Going forward, that is the model used for the \textit{Attack Surface Detection Engine}

\subsubsection{RQ2: How effectively does SIExVulTS detect CWE-200 vulnerabilities in benchmark scenarios?}

\begin{table}[h]
\centering
\caption{Performance of the proposed tool on the benchmark dataset.}
\label{tbl:bertonToy}
\begin{tabular}{|c|c|c|c|c|c|c|}
\hline
\textbf{CWE} & \textbf{GOOD} & \textbf{BAD} & \textbf{Total} & \textbf{Precision} & \textbf{Recall} & \textbf{F1 Score} \\
\hline
201 & 7  & 10 & 17 & 100.00  & 80.00  & 88.89 \\
\hline
204 & 11 & 11 & 22 & 100.00  & 90.91 & 95.24 \\
\hline
208 & 10 & 13 & 23 & 92.86  & 100.00  & 96.30 \\
\hline
209 & 17 & 17 & 34 & 100.00  & 76.47  & 86.67 \\
\hline
214 & 11 & 10 & 21 & 100.00  & 80.00  & 88.89 \\
\hline
215 & 10 & 10 & 20 & 100.00  & 70.00  & 82.35 \\
\hline
532 & 10 & 13 & 23 & 100.00 & 84.62 & 91.67 \\
\hline
535 & 10 & 15 & 25 & 100.00 & 40.00 & 57.14 \\
\hline
536 & 10 & 10 & 20 & 100.00  & 80.00  & 88.89 \\
\hline
537 & 10 & 10 & 20 & 100.00  & 80.00 & 88.89 \\
\hline
538 & 3 & 2 & 5 & 100.00 & 100.00 & 100.00 \\
\hline
550 & 10 & 10 & 20 & 100.00  & 80.00  & 88.89 \\
\hline
598 & 10 & 10 & 20 & 100.00  & 70.00  & 82.35 \\
\hline
615 & 10 & 10 & 20 & 100.00  & 50.00  & 66.67 \\
\hline
\textbf{Total} & \textbf{139} & \textbf{161} & \textbf{300} & \textbf{99.13} & \textbf{75.50} & \textbf{85.71} \\
\hline
\end{tabular}
\end{table}
We evaluated SIExVulTS on our benchmark dataset, which includes samples labeled across various CWE-200 subtypes. As shown in Table~\ref{tbl:bertonToy}, the tool achieves an overall F1 score of 85.71\%.

The lowest performance was observed for CWE-535, which involves information exposure through shell error messages. This category typically depends on complex runtime conditions that are difficult to capture statically. CWE-615 (sensitive comments) also showed relatively low performance, primarily because it does not involve dataflow. As a result, both the Exposure Analysis Engine and the Flow Verification Engine are bypassed, and detection relies solely on the Attack Surface Detection Engine, which may reduce effectiveness.

\subsection{RQ3: How effective is SIExVulTS in real-world CWE-200 detection?}
To validate the effectiveness and efficiency of SIExVulTS, we applied it to real-world open-source projects. To identify suitable projects for analysis, we developed a script to locate reported CVEs related to CWE-200. This process was refined to focus on open-source Java projects. Among the identified candidates, Jenkins was selected due to its multiple CWE-200-related CVEs and suitability for analysis based on clear vulnerability references in its source code.

We collected Jenkins CVEs from GitHub's advisory database and the NVD. Of the 15 CVEs with available vulnerable versions, SIExVulTS successfully detected 9. Table~\ref{tbl:jenkinsCVEs} summarizes the results. Missed cases largely stemmed from unrecognized sink types, particularly related to REST API usage. Improving the sink model to recognize such patterns is a future work direction.

\begin{table}[h]
\centering
\caption{SIExVulTS evaluation on reported vulnerabilities of Jenkins \label{tbl:jenkinsCVEs}}
\label{tbl:jenkinsCVEs}
\begin{tabular}{|l|c|c|c|} 
\hline
\textbf{CVE-ID}           & \textbf{CWE} & \textbf{Version} & \textbf{Detected}   \\ 
\hline
% \hline
CVE-2018-1000169 & 200    & 2.107.1      & YES          \\ 
\hline
CVE-2018-1000192 & 200    & 2.107.1      & YES            \\ 
\hline
CVE-2017-1000395 & 200    & 2.73.1       & NO               \\ 
\hline
CVE-2017-1000398 & 200    & 2.73.1       & NO              \\ 
\hline
CVE-2017-1000399 & 200    & 2.73.1       & NO               \\ 
\hline
CVE-2020-2102    & 208    & 2.218        & YES             \\ 
\hline
CVE-2016-0790    & 208    & 1.649        & YES            \\ 
\hline
CVE-2016-0791    & 200    & 1.649        & YES             \\ 
\hline
CVE-2018-1000862 & 200    & 2.138.1      & YES             \\ 
\hline
CVE-2018-1000410 & 200    & 2.138.1      & NO               \\
\hline
CVE-2014-3662    & 200    & 1.582        & YES             \\ 
\hline
CVE-2017-2606    & 200    & 2.32.1       & NO               \\ 
\hline
CVE-2014-2064    & 200    & 1.550         & YES            \\ 
\hline
CVE-2018-1999006 & 200    & 2.121.1      & NO               \\ 
\hline
CVE-2022-34174   & 208    & 2.355        & YES             \\
\hline
\end{tabular}
\end{table}

\subsubsection{Precision Analysis With and Without Flow Verification}

A key goal of SIExVulTS is to ensure that the detections are accurate and reliable. We performed a simple study to test the effect of our Flow Verification Engine. Table~\ref{tbl:precisionimpact} shows a comparison of the tool's precision with and without this engine. The verification step significantly improves overall performance by filtering out invalid flows flagged by static analysis alone. Bringing the global precision without verification from 22.6\% to 87.2\% across 31 projects tested, 28 of them saw an improvement in their precision, while 3 remained the same, and none of them reduced. 

\begin{table}[h]
\centering
\caption{Impact of Flow Verification on Evaluation Metrics}
\label{tbl:precisionimpact}
\begin{tabular}{|l|c|c|}
\hline
\textbf{Metric} & \textbf{Without Verification} & \textbf{With Verification} \\
\hline
Precision & 0.226 & \cellcolor{gray!20}\textbf{0.872} \\
\hline
Recall    & \cellcolor{gray!20}\textbf{1.000} & 0.938 \\
\hline
F1 Score  & 0.369 & \cellcolor{gray!20}\textbf{0.904} \\
\hline
Accuracy  & 0.226 & \cellcolor{gray!20}\textbf{0.955} \\
\hline
\end{tabular}
\end{table}

\subsection{Discovery of New CVEs}

In addition to evaluating known vulnerabilities, SIExVulTS was used to proactively audit real-world open-source projects for instances of sensitive information exposure. Candidate projects were selected based on relevance to the Java/Maven ecosystem and prior security disclosures. SIExVulTS analyzed each target project for sensitive dataflows, and its outputs were manually reviewed to validate whether the findings represented genuine vulnerabilities. In promising cases, proof-of-concept (PoC) exploits were constructed to demonstrate exploitability.

As of this writing, 6 previously unknown vulnerabilities have been confirmed and assigned CVE identifiers, with 12 additional reports currently under review by the respective project security teams.

\begin{itemize}
  \item \textbf{CVE-2025-26795} – \textit{Apache IoTDB (v1.3.3)}: JDBC driver logs plaintext database credentials.
  
  \item \textbf{CVE-2025-26864} – \textit{Apache IoTDB (v1.3.3)}: OpenID authentication module logs user passwords during failed login attempts.
  
  \item \textbf{CVE-2025-30677} – \textit{Apache Pulsar (v4.0.2 \& v4.0.3)}: Kafka connectors log sensitive configuration properties, including SSL truststore passwords.
  
  \item \textbf{CVE-2025-53651} – \textit{Jenkins HTML Publisher Plugin (v4.2.5)}: Logs absolute file paths, potentially exposing system-level information.
  
  \item \textbf{CVE-2025-48955} – \textit{Para (v1.50.6)}: Logs AWS `accessKey` and `secretKey` at the INFO level during application health checks.
  
  \item \textbf{CVE-2025-49009} – \textit{Para (v1.50.6)}: Logs Facebook OAuth `accessToken` at the WARN level during authentication failures.
\end{itemize}

These findings demonstrate SIExVulTS’s ability to identify impactful, real-world vulnerabilities that were previously unknown, validating the tool’s relevance beyond synthetic or benchmark datasets.

%% file: Discussion.tex
\section{Discussion}
\label{sec:discussion}
\subsection{Results}
We proposed a novel system, SIExVulTS, to address the underrepresented challenge of detecting CWE-200 vulnerabilities. Our approach integrates transformer-based embeddings with static analysis to identify sensitive information and its potential exposure points, tracing how this information propagates through source code paths. The system is composed of three core components: the 	
\textit{Attack Surface Detection Engine}, which detects sensitive sources and sinks; the 	\textit{Exposure Analysis Engine}, which uses custom CodeQL queries to map potential dataflows to the CWE-200 hierarchical model based on the CWE repository; and the \textit{Flow Verification Engine}, which evaluates the semantic validity of flows to eliminate false positives.

\subsubsection{Attack Surface Detection Engine}
To evaluate the effectiveness of our \textit{Attack Surface Detection Engine}, we compiled a comprehensive dataset of existing CVEs related to CWE-200 and conducted a series of experiments. The results indicate that our approach detects sensitive sources and sinks within source code with an average F1 score greater than 93\%. The \textit{Attack Surface Detection Engine}, underpinned by SentBERT, was the most effective at this task compared to ChatGPT-4o, CodeBERT, and CodeT5.

\noindent Several factors contribute to these observations:

\begin{enumerate}
\item The task of finding sensitive information is inherently broader for ChatGPT-4o compared to the encoder-based approaches of the other three models. ChatGPT-4o is tasked with identifying all occurrences of a specific sensitive information category within a given file. In contrast, encoder-based models enable more targeted queries by allowing for the direct identification of each particular occurrence. This targeted approach is challenging to replicate with ChatGPT-4o because the number of required queries would escalate significantly—from a few thousand (on a per-file basis) to potentially millions (on a per-category basis). This dramatic increase in queries not only complicates the process but also incurs substantially higher computational costs.

\item While ChatGPT-4o benefits from an extended context window of 128k tokens, our experiments revealed that its performance degrades when processing inputs exceeding approximately 5k tokens. Specifically, the model begins to overlook or fail to classify portions of the data, which diminishes its accuracy in identifying sensitive information.

\item In addition, we observed that ChatGPT-4o sometimes struggles to differentiate between the different categories. For example, it is confusing to detect string literals and comments numerous times. This led to skipping data classification, making it unreliable and difficult to reproduce. This can be seen in Table \ref{tbl:performance_comparison}, where ChatGPT-4o performs much worse on comments compared to the other categories. Efforts were made to counteract this, such as injecting the parsed data into the prompts, so that the model wouldn't have to perform the tasks of parsing the data and then classifying it. This had ~10\% improvement in the F1 score, but still didn't completely eliminate the issue of missing data. 
\end{enumerate}

\subsubsection{Exposure Analysis Engine}
To evaluate the effectiveness of our Exposure Analysis Engine, we conducted a series of experiments using our benchmark dataset. This component was designed to trace the flow of sensitive data from sources to sinks using custom CodeQL queries aligned with the CWE-200 hierarchy. The results show that the engine was able to identify valid detections across a wide range of CWE-200 subcategories.

Contributing factors to these results include:
\begin{enumerate}
    \item By instantiating queries only for previously identified sources and sinks, the engine achieved high efficiency, avoiding exhaustive or irrelevant code scanning.
    \item The design of queries incorporated semantic understanding of typical vulnerability patterns, which helped filter out spurious paths and increased the accuracy of flow detection.
\end{enumerate}

\subsubsection{Flow Verification Engine} To evaluate the effectiveness of the Flow Verification Engine, we tested its ability to validate source-to-sink flows identified by the Exposure Analysis Engine. This component operates on CodeQL-generated detections and independently verifies each dataflow by reconstructing a semantically rich representation that includes the surrounding source code context. These reconstructed flows are embedded using GraphCodeBERT and processed with a Transformer-based aggregator to capture long-range dependencies and structural patterns.

The results demonstrate that the Flow Verification Engine substantially improves precision, particularly for CWEs involving complex or noisy flows, such as CWE-209 and CWE-204. It serves as a critical final validation step that filters out false positives from static analysis. Particularly ones where sensitive data flows into complex objects, such as maps, and another part of the object is actually exposed.  

Contributing factors to these results include:
\begin{enumerate}
\item Parsing the source code around each flow step allowed the engine to analyze real program context.
\item GraphCodeBERT embeddings combined with Transformer-based aggregation enabled effective reasoning over long or deeply nested flows.
\item Performing this validation after CodeQL flow detection created a layered analysis pipeline, enhancing accuracy without compromising performance.
\end{enumerate}

\subsection{Limitations}
While SIExVulTS demonstrates strong performance across a range of CWE-200 subcategories, several limitations remain. First, although dynamic sink detection expands the range of identifiable exposure points, it still misses nuanced cases. This was evident in our evaluation of Jenkins CVEs, where vulnerabilities involving REST API disclosures (e.g., CVE-2017-2606) were not detected. These patterns often fall outside the predefined sink heuristics and require further training data and refinement of the sink detection model.

Second, the system is inherently limited in its ability to detect vulnerabilities dependent on runtime behavior—such as access control violations, user authentication flows, or permission checks. These aspects are typically beyond the scope of static analysis and requires execution context to resolve accurately.

Finally, while transformer models improved the semantic understanding of sensitive components, they may still misclassify borderline cases due to the subjectivity of what constitutes “sensitive” data across different domains.

To address these issues, future work will focus on expanding the training datasets (particularly with REST-related patterns) and refining sink heuristics. These additions will improve both the completeness and precision of CWE-200 vulnerability detection in real-world systems.

\subsection{Comparison to Existing Tools} We attempted to compare our tool's performance to existing tools such as DeepWukong \cite{cheng2021deepwukong},  VulDeePecker \cite{li2018vuldeepecker}, and commercial tools such as SonarQube \cite{sonarqube}. However, we were unable to find any tools that specifically covered CWE-200 for Java. As a test, we attempted to use SonarQube with Jenkins version 2.107.1 and were unable to find any detections related to CWE-200. Since it does not officially look for CWE-200 vulnerabilities, we did not proceed with all the versions in Table \ref{tbl:jenkinsCVEs}. The closest tool we found was LineVul \cite{fu2022linevul}, which covers C/C++. While this does not allow us to perform a direct comparison, it does demonstrate a gap in the current coverage of tools regarding CWE-200. Which is especially concerning given its importance \cite{iex1}.

\subsection{Threat to Validity}
\subsubsection{Threats to Internal Validity} 

\textbf{Dataset Quality}:
One threat to internal validity is related to the correctness and completeness of the benchmark dataset. Given that the authors developed the benchmark dataset, it may not cover all possible code examples for each CWE, and defining the vulnerable code samples is error-prone. To mitigate this, we cross-checked the code samples and revised them based on author agreements. In addition, we collected real data from the NVD repository to mitigate this threat. These real samples were collected from projects with different sizes, complexity, and usages, all of which contain CWE-200 vulnerabilities.

\textbf{Labeling Errors}:
Another threat to the validity of our approach is the labeling of data as sensitive or non-sensitive and API calls as sinks or not. This large manual process is prone to mistakes, and the data sensitivity can change based on the context. We tried to mitigate this issue in two ways:
\begin{enumerate}
\item Considering the context and application of the open-source project during manual review 
\item Peer-reviewing the labeled data by the authors and having discussions on disagreements. The Cohen’s Kappa coefficients are shown in Table~\ref{tbl:cohen_kappa}. Overall, reviewers demonstrated strong agreement on the classifications, with variables showing the lowest agreement, as expected, given the subjectivity involved in classifying variables.

\end{enumerate}

\begin{table}[h]
  \centering
  \caption{Cohen’s Kappa Coefficients for Agreement by Category}
  \label{tbl:cohen_kappa}
  \begin{tabular}{|l|c|}
    \hline
    \textbf{Category} & \textbf{Coefficient} \\ \hline
    Variables         & 0.68                  \\ \hline
    Strings           & 0.85                  \\ \hline
    Comments          & 0.83                  \\ \hline
    Overall           & 0.77                  \\ \hline
  \end{tabular}
\end{table}

\subsubsection{Threats to External Validity} We evaluated the model based on our collected data, which can affect the model's performance. To mitigate this threat, we collected real data from projects of different sizes, complexities, and usage contexts to make the model more generalizable. Additionally, we evaluated the tool's performance in addressing real vulnerabilities reported in Jenkins. 

\subsubsection{Threats to Construct Validity} One significant challenge in this study is the inherent subjectivity in defining what constitutes sensitive information. The perception of sensitivity can vary across individuals, contexts, and domains, making it difficult to establish a universally accepted definition. For instance, while certain configuration details or comments may be deemed sensitive in one application, they might be considered non-sensitive in another. This variability introduces potential ambiguity in labeling and evaluation. To mitigate this, we based our definitions on standardized guidelines from the CWE-200 hierarchical model and conducted peer reviews during the dataset creation process to ensure consistency and reduce bias.

%% file: Ethics.tex
\section{Ethics and Open Science}

This study follows ethical research practices and supports open science. All datasets were either publicly available or synthetically generated to reflect realistic scenarios without containing private or personally identifiable information (PII).

We recognize the potential misuse of tools like SIExVulTS and stress that it is intended solely to support secure software development. Its goal is to help identify and mitigate sensitive data exposures—not to enable exploitation.

To promote transparency and reproducibility, the source code, CodeQL queries, and synthetic datasets are available in a publicly accessible repository \cite{hawaii-pique-cwe200}.

%% file: Conclusion.tex
\section{Conclusion}
\label{sec:conclusions}
We presented \textbf{SIExVulTS}, a novel vulnerability detection system that targets the under-addressed CWE-200 category of sensitive information exposure (SIEx) in Java applications. By integrating transformer-based embeddings from SentBERT with static analysis via CodeQL, SIExVulTS detects both the presence and propagation of sensitive data with high precision. The system’s three-stage architecture—Attack Surface Detection, Exposure Analysis, and Flow Verification—enables end-to-end detection of diverse SIEx vulnerabilities. Experimental results across benchmark datasets, real-world CVEs, and manually labeled dataflows demonstrate the system's strong performance, achieving an F1 score above 93\% for detecting sensitive elements and 85.71\% for identifying CWE-200 vulnerabilities. In addition, SIExVulTS uncovered multiple previously undocumented vulnerabilities in major open-source projects, further validating its practical utility. Future work will explore the integration of additional domain-specific heuristics to improve recall and support broader language ecosystems.

% \section{Data Availability}
% We will make data publicly available upon acceptance.

%% file: Acknowledgement.tex
\section*{Acknowledgments}
This research was conducted with support from the U.S. Department of Homeland Security (DHS) Science and Technology Directorate (S\&T) under contract 70RSAT22CB0000005. Any opinions contained herein are those of the author and do not necessarily reflect those of DHS S\&T.

%% file: Appendix.tex
\section{Appendix}
\begin{figure*}[t]
\centering
\begin{tcolorbox}[width=0.90\linewidth, boxrule=0.5mm, colframe=black!75!black, colback=black!10!white, enhanced, title=ChatGPT Prompt for Identifying Sensitive Variables]
\scriptsize

\#\#\# Task
I want you to detect sensitive variables in the Java Source Code.
Sensitive variables are those that, if exposed, could lead to a vulnerability or compromise the security and integrity of a system.

\#\#\# Sensitive Variable Categories:
\begin{enumerate}
    
 \item Authentication and Authorization and Credentials Information: Variables holding passwords, API keys, keys, usernames, verification codes, credentials, etc.
 \item Personal Identifiable Information (PII): Variables containing names, emails, addresses, social security numbers, health information, accounts, etc.
 \item Financial Information: Variables related to credit cards, bank account numbers, account IDs, payment IDs, CVV, etc
 \item Files Containing Sensitive Information, Sensitive File Paths, URLs/URIs: Variables storing internal URLs/URIs or file paths to files that contain sensitive information or files themselves.
 \item Sensitive System and Configuration Information: Variables with database, cloud provider, or network connection strings, database schemas, configuration details, environment variables, sensitive settings, controllers, and managers.
 \item Security and Encryption Information: Variables holding encryption keys, seeds, or certificates.
 \item Application-Specific Sensitive Data: Variables storing sensitive information such as device details (Names, IDs, properties, objects), Application-specific IDs, email messages, notifications, etc.
 \item Query Parameters: Variables storing sensitive data in HTTP GET requests.
\end{enumerate}
\#\#\# Note
Exclude variables related to handlers, wrappers, loggers, listeners, generic file paths, and URLs.

\#\#\# File Markers
Each file begins with "-----BEGIN FILE: [FileName]-----" and ends with "-----END FILE: [FileName]-----".

\#\#\# Report Format
Provide a JSON response in the following format. Do not include any error messages or notes:
\begin{enumerate}
\item Where it says "variableName1" and "variableDescription1" you should replace them with the actual name and description of the sensitive variable.
\item Provide a JSON response for each file that matches the format below. 
\begin{itemize}
  \item The "name" field should be the sensitive information found in the variable.
  \item The "description" field should tell which category the variable belongs to.
\end{itemize}
\item Make sure there are no duplicate entries in the response.
\end{enumerate}
\begin{verbatim}
{   "files": [
        {"fileName": "FileName1.java",
          "sensitiveVariables": [
            {   "name": "variableName1",
                "description": "variableDescription1"},
            {   "name": "variableName2",
                "description": "variableDescription2"}]
        }]}
\end{verbatim}
\end{tcolorbox}
\vspace{-8pt}
\caption{Prompt asks ChatGPT to identify sensitive variables in source code}
\label{fig:prompt}
\end{figure*}

\begin{figure*}[h]
\centering
\begin{tcolorbox}[width=0.90\linewidth, boxrule=0.5mm, colframe=black!75!black, colback=black!10!white, enhanced, title= Simplified CodeQL Query for CWE-537]
\scriptsize
\begin{verbatim}
/**
 * @name CWE-537: Exposure of sensitive information in runtime error messages
 * @kind path-problem
 */
 import java
 import semmle.code.java.dataflow.TaintTracking
 import semmle.code.java.dataflow.FlowSources
 import CommonSinks.CommonSinks
 import SensitiveInfo.SensitiveInfo
 module Flow = TaintTracking::Global<RuntimeSensitiveInfoExposureConfig>;
 import Flow::PathGraph
 
 module RuntimeSensitiveInfoExposureConfig implements DataFlow::ConfigSig {
  predicate isSource(DataFlow::Node source) {
    exists(SensitiveVariableExpr sve |
      source.asExpr() = sve <- Input from Attack Surface Detection Engine)}
 
   predicate isSink(DataFlow::Node sink) {
    // Sink exposes sensitive info inside a RuntimeException catch block

     exists(MethodCall mc, CatchClause cc |
       cc.getACaughtType().getASupertype*().hasQualifiedName("java.lang", "RuntimeException") and
       mc.getEnclosingStmt().getEnclosingStmt*() = cc.getBlock() and
       (
         CommonSinks::isPrintSink(sink) or
         CommonSinks::isErrorSink(sink) or
         CommonSinks::isServletSink(sink) or
        getSinkAny(sink) <- Attack Surface Sinks
       ) and
       sink.asExpr() = mc.getAnArgument())}}

 from Flow::PathNode source, Flow::PathNode sink
 where Flow::flowPath(source, sink)
 select sink.getNode(), source, sink, "Java runtime error message containing sensitive information"
\end{verbatim}
\end{tcolorbox}
\vspace{-8pt}
\caption{CodeQL query to detect sensitive data in exception messages (CWE-537)}
\label{fig:codeql537}
\end{figure*}